\documentclass{article}

\usepackage{arxiv}

\usepackage[utf8]{inputenc} 
\usepackage[T1]{fontenc}    
\usepackage{hyperref}       
\usepackage{url}            
\usepackage{booktabs}       
\usepackage{amsfonts}       
\usepackage{nicefrac}       
\usepackage{microtype}      
\usepackage{multirow}
\usepackage{graphicx}
\usepackage{natbib}
\usepackage{doi}
\usepackage{float}
\usepackage{pdfpages}

\title{Detection Threshold of Audio Haptic Asynchrony in a Driving Context}

\author{ \href{https://orcid.org/0000-0003-3704-1415}{\includegraphics[scale=0.06]{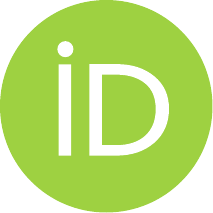}\hspace{1mm}Gyanendra Sharma} \\
	Toyota Research Institute\\
	Los Altos, CA 94022 \\
	\texttt{gyanendra.sharma870@gmail.com} \\
	\And
	\href{https://orcid.org/0000-0003-2708-8623}{\includegraphics[scale=0.06]{pics/orcid.pdf}\hspace{1mm}Hiroshi Yasuda} \\
	Toyota Research Institute\\
	Los Altos, CA 94022 \\
	\texttt{hiroshi.yasuda@tri.global} \\
	\And
        \href{https://orcid.org/0000-0002-2852-4020}{\includegraphics[scale=0.06]{pics/orcid.pdf}\hspace{1mm}Manuel Kuehner} \\
	Toyota Research Institute\\
	Los Altos, CA 94022 \\
	\texttt{manuel.kuehner@gmail.com}
}

\date{}

\renewcommand{\headeright}{ }
\renewcommand{\undertitle}{ }
\renewcommand{\shorttitle}{Haptic Audio Asynchrony Detection Threshold}

\hypersetup{
pdftitle={Detection Threshold of Audio Haptic Asynchrony in a Driving Context},
pdfauthor={Gyanendra Sharma, Hiroshi Yasuda and Manuel Kuehner},
pdfkeywords={Multimodal signals, Psychophysics, Detection Threshold, Driving, Haptic Feedback},
}

\begin{document}

\maketitle

\begin{abstract}
In order to provide perceptually accurate multimodal feedback during driving situations, it is vital to understand the threshold at which drivers are able to recognize asyncrony between multiple incoming Stimuli. In this work, we investigated and report the \textit{detection threshold} (DT) of asynchrony between audio and haptic feedback, in the context of a force feedback steering wheel. We designed the experiment to loosely resemble a driving situation where the haptic feedback was provided through a steering wheel (\textit{Sensodrive}), while the accompanying audio was played through noise cancelling headphones. Both feedbacks were designed to resemble rumble strips, that are generally installed on the side of major roadways as a safety tool. The results indicate that, for $50\%$ of the participants, asynchrony was detectable outside the range of -75 ms and 110 ms, where the former is related to perceiving audio before haptic and vice versa for the latter. We were also able to concur with previous studies, which state that latency is perceivable at a lower threshold when audio precedes haptic stimuli. 
\end{abstract}
\section{Introduction}
Rumble strips on the side of the road are a major safety tool to make sure drivers do not drift off the road during driving situations. These physical artifacts assert driver's attention by providing both auditory and haptic feedback. Designing safety systems that translate similar phenomena in a synthetic manner,  so that drivers can be warned in cases of road departures holds significant promise. Specifically, an interaction concept that offers digitally created rumble strip phenomena through audio and haptic feedback can be created to warn drivers when they unintentionally depart the lane or the road itself. One key advantage of such a concept is an easy translation of mental model from an already existing warning mechanism that is widely understood and familiar, especially in North America.

\begin{figure}[h!]
\centering
\includegraphics[width = 0.25\linewidth]{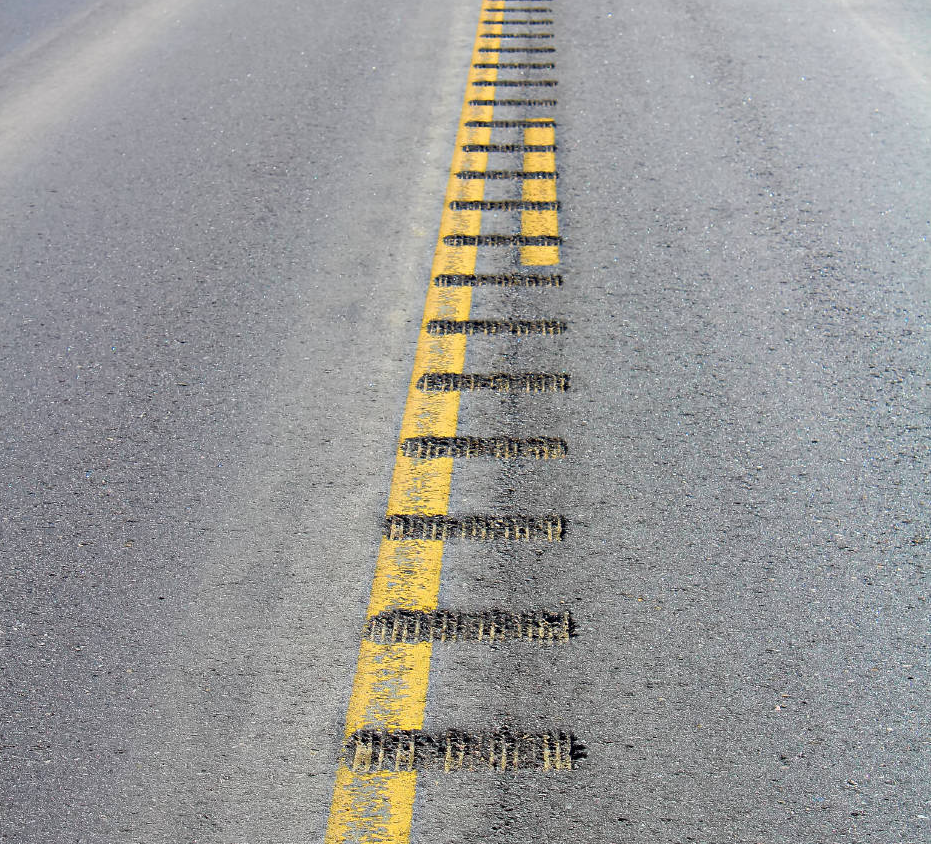}
\caption{Rumble Strip in a physical environment.}
\label{fig:rumble}
\end{figure}
\begin{figure}
\centering
\includegraphics[width = 0.99\linewidth]{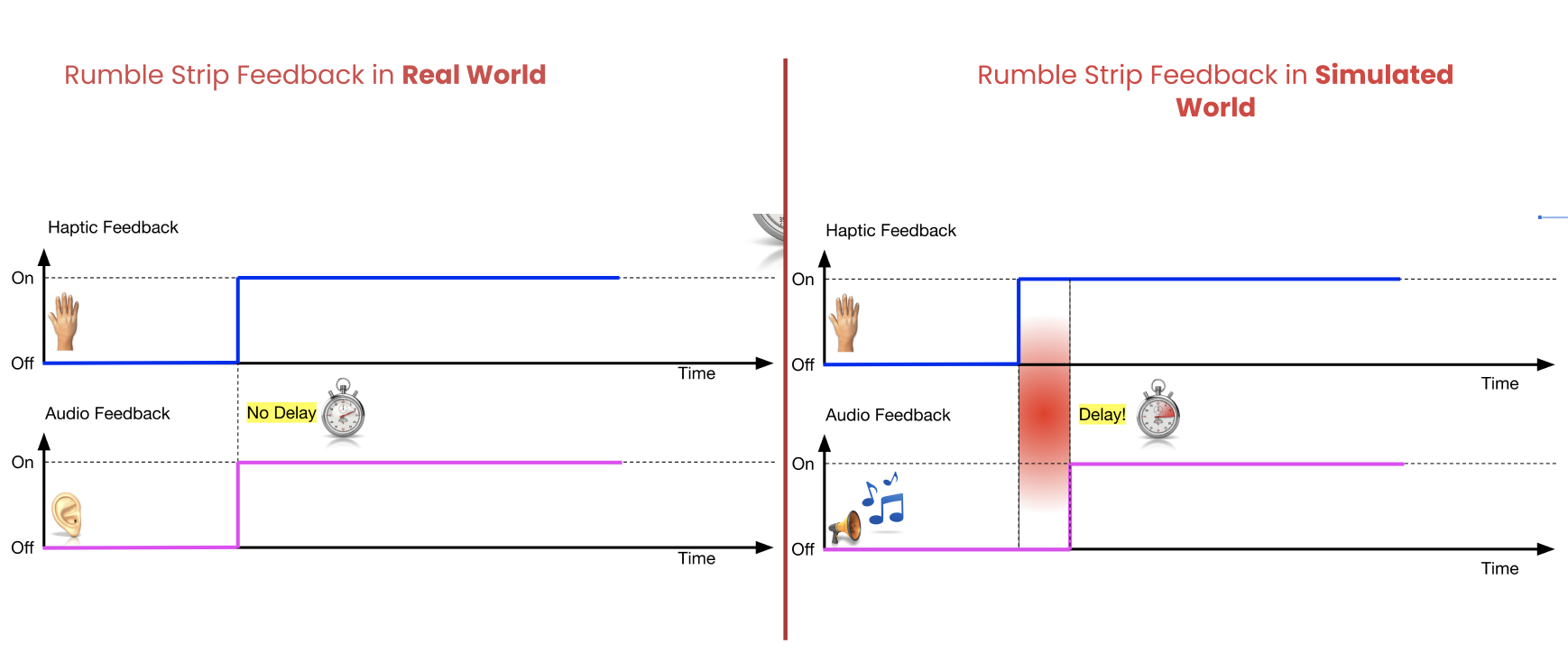}
\caption{Latency between audio and haptic stimuli in the context of real world vs synthetic rumble strips.}
\label{fig:timing}
\end{figure}

In order to do this, there are various parameters that characterize the feedback, and one of the major components is the timing at which both audio and haptic responses are delivered to the driver. While physical rumble strips, as shown in figure \ref{fig:rumble} are characterized by natural law of physics, synthetic or electronically created feedback systems require each signal to propagate through a separate set of actuators, control units and computation units before they arrive to the driver. This causes delay in regards to the delivery of either of the haptic or audio stimuli to the driver. Figure \ref{fig:timing} shows the case where haptic stimuli is delivered to the driver slightly later than audio. A large asynchrony between the stimuli representing the same feedback event can lead to confusing user feedback systems. So, for these feedback systems to be effective and also to avoid annoyance, which has been shown to be a major reason for safety systems being turned off,  it is pertinent that both audio and haptic signal onset time are perceptually in sync, even if asynchrony exists in the underlying system. The detection threshold (DT) at which this asynchrony is perceivable to the driver is the major contribution of this work.

Human beings can only detect asynchrony outside of a certain time window. This is also referred to as the point of subjective simultaneity (PSS), where stimulus onset asynchrony (SOA) are most likely to yield responses that affirm that the SOA was detectable \cite{stone2001now}. Early works on this topic are derived from the field of psychology, where SOA between audio and visual stimuli were studied \cite{hershenson1962reaction, woodworth1954experimental}. More recently, haptic has also been included to study the perception of cross-modal simultaneity \cite{adelstein2003sensitivity, levitin2000perception, altinsoy2003perceptual}. Beyond the reported DT numbers, these studies also demonstrated that latency is perceived at a lower threshold when audio precedes the haptic feedback.

While earlier studies focused on a more physical setup, recent studies have investigates this topic in the context of modern applications such as specific computer interfaces, touchscreens and force feedback haptic devices. For instance, work of Silva et al. \cite{silva2013human} was specifically focused on computer based multimedia application and showed that asynchrony of audio played before haptic within 92 ms and, audio played after haptic within 110 ms was not detected by the participants. With the rapid increase in usage of AR and VR technologies in recent years, the application areas upon which this topic is being investigated has expanded significantly. Tele-operation \cite{neumeier2019teleoperation, liu2017investigating}, multi-sensory media (mulsemedia) \cite{covaci2018multimedia, yuan2015perceived} are a few examples. 

Through this user study, we investigate the detection threshold (DT) at which drivers are able to perceive asynchrony between haptic and audio. The DT is determined using a 3 alternative forced choice (3AFC) adaptive staircase procedure \cite{prins2016psychophysics, gescheider1997psychophysics}. Threshold obtained in this manner corresponds to the 50 percentile point on a psychometric function \cite{wichmann2001psychometric}.

In particular, we are interested in the following research questions.
\begin{itemize}
    \item What is the time difference at which drivers begin to notice asynchrony between haptic and audio feedback.
    \item Is the previous conclusion, \textit{latency is perceived at a lower threshold when audio precedes the haptic}, applicable to driving context?
\end{itemize}

\section{System Characterization}
\begin{figure}
\centering
\includegraphics[width = 0.55\linewidth]{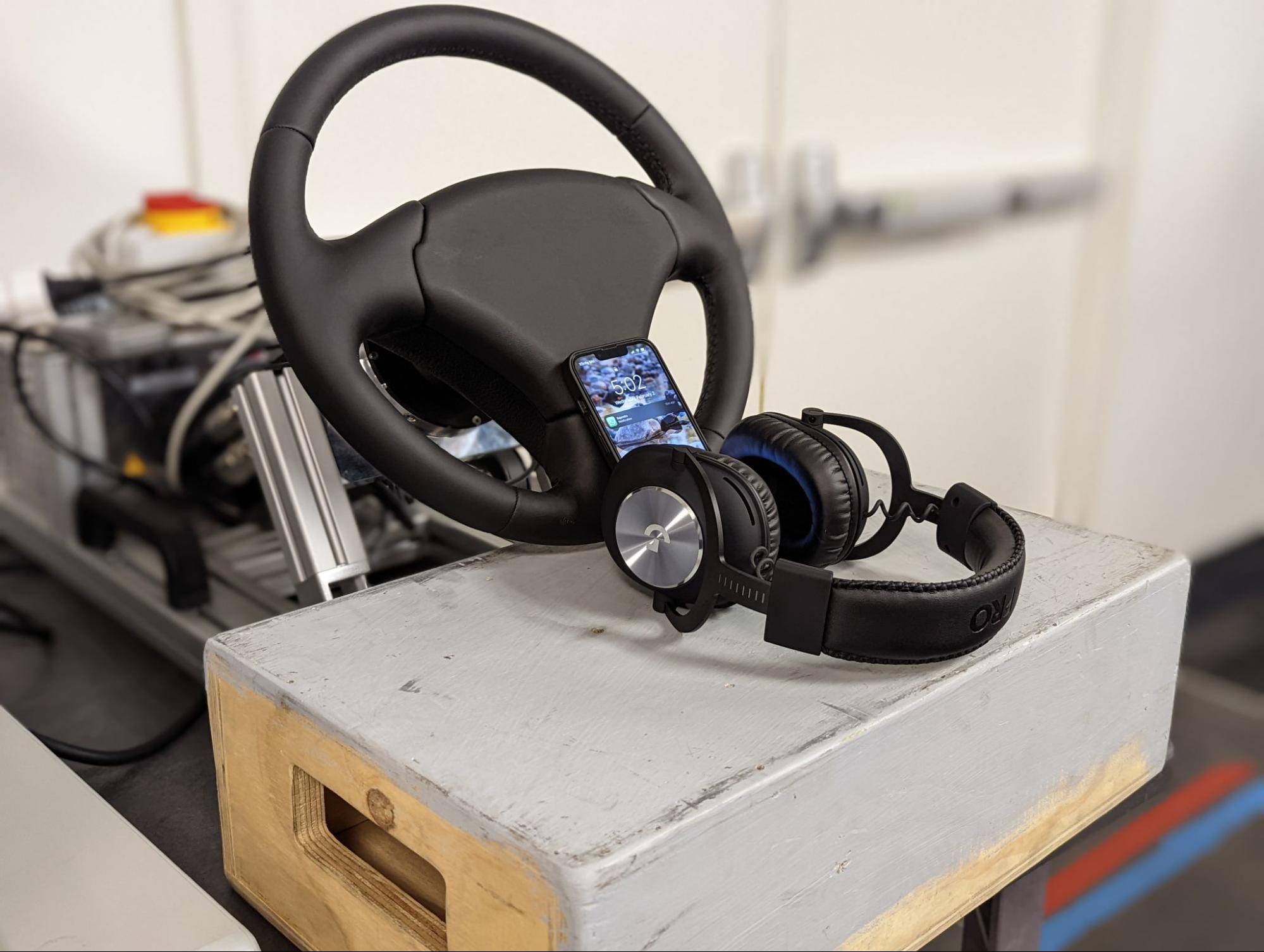}
\caption[]{Setup to record audio and vibration.}
\label{fig:system}
\end{figure}

Before embarking on designing a system that is able to accurately provide desired offsets between audio and haptic feedback, we had to explore the role of system latency i.e the inherent latency within the computing environment of the workstation. We investigated this by recording repeated samples of audio-haptic feedback pairs, as shown in figure \ref{fig:system}, and noting the time difference through analysis of their respective frequencies.

\begin{figure}
\centering
\includegraphics[width = 0.45\linewidth]{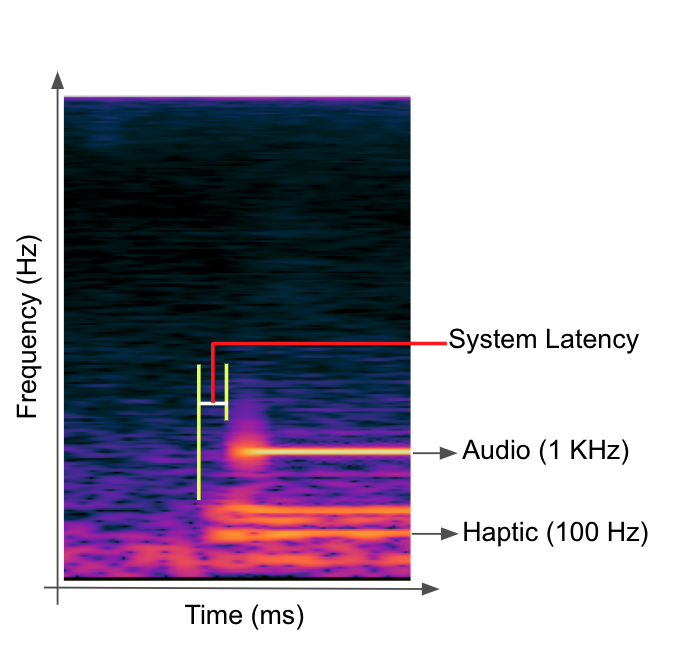}
\caption[]{Spectrogram to analyze the offset between audio and haptic.}
\label{fig:spectrogram}
\end{figure}

We used a total of 58 audio-haptic pairs and analyzed each pair to determine their actual offset time and compared it with the desired offset time to obtain a value for system latency. The audio had variety of offsets ranging between 6 ms to 30 ms and it was played at 1KHz while the vibration was played at 100 Hz. As shown in figure \ref{fig:spectrogram}, a spectrogram for each feedback pair was analyzed to determine the actual offset. The audio onset time ($ao$), haptic onset time ($ho$), desired offset ($do$) from all samples ($N = 58$) was used to calculate the system latency ($sl$) as follows.
\begin{equation}
    sl = \frac{1}{N}{\sum_{\alpha = 1}^{N} ((ao_{\alpha} - ho_{\alpha}) - do_{\alpha})}
    \label{eqn:system_latency}
\end{equation}

Using equation \ref{eqn:system_latency}, we calculated the system latency on average to be +15.93 ms, with median value of 16 ms and standard deviation of 4.61 ms. This showed that audio on average gets delayed by approximately 15.93 ms due to system latency even in cases when both are played simultaneously. We included the finding in our experimental design and setup to compensate for this system delay. 
\section{Methodology}
In this study, participants held onto a Sensodrive steering wheel \cite{sensodrive} that was connected to a Linux workstation while wearing a pair of noise canceling wired headphones. Participants were exposed to 3 sets of haptic and audio feedback pairs with 3 seconds interval between each set. The haptic feedback appeared in the form of vibration on the Sensodrive wheel while the audio feedback was played through the headphone. Two of the audio-haptic feedback pairs were in sync while the third one was played out of sync with experimentally controlled offset values. The order at which each of these sets appeared was randomized. The pattern of the feedback for both the sound and haptic were designed to resemble a rumble strip. After each trial, participants were asked the following question to understand their perception of the feedback;  “which audio-haptic pair do you think was NOT in sync?”

\begin{figure}
\centering
\includegraphics[width = 0.95\linewidth]{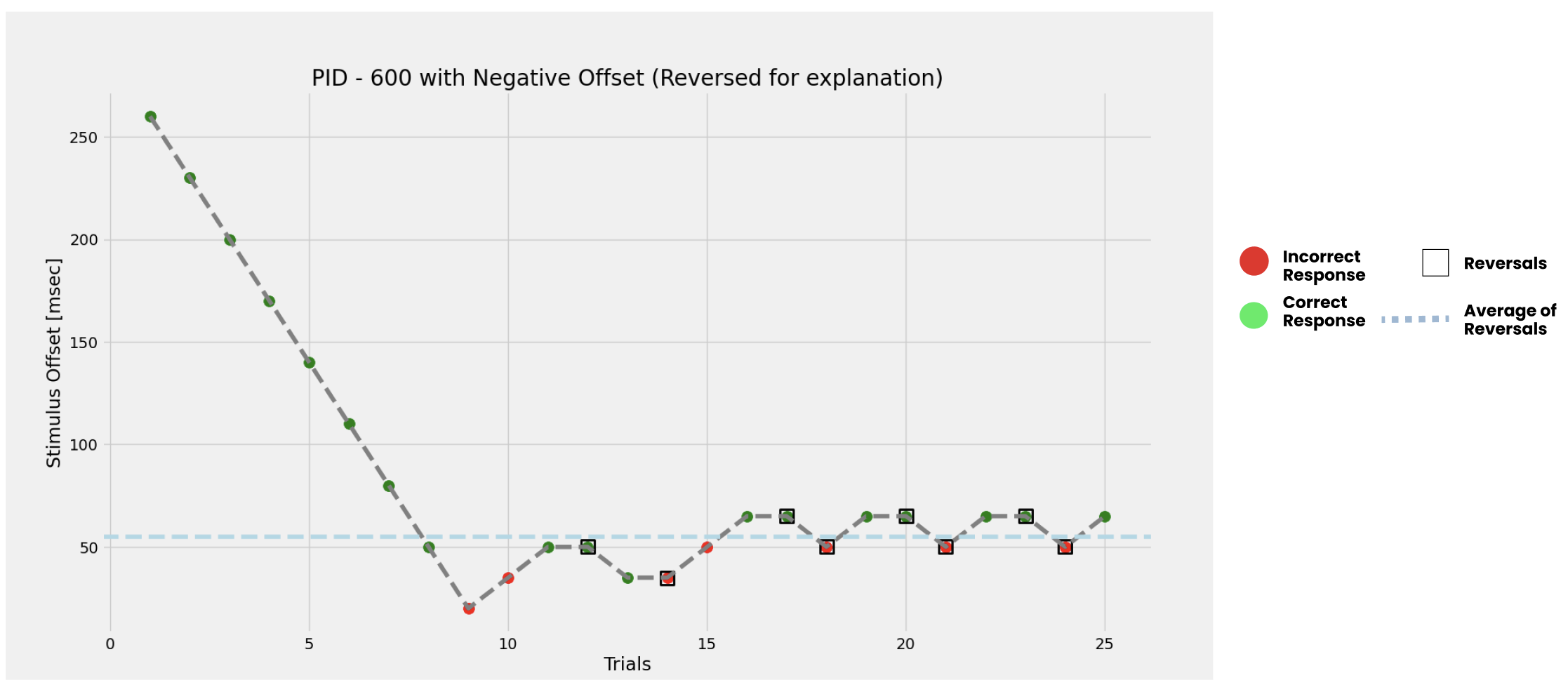}
\caption{Sample participant response based on the 2 step up, 1 step down staircase approach.}
\label{fig:user_response}
\end{figure}

We employed a 3AFC (3 alternatives force choice) method i.e participants can respond with only one of the three given choices. In this particular case, they were instructed to respond by speaking out one of the three numbers; 1, 2 or 3 based on whether the first, second or third feedback pair was out of sync. Their responses were recorded by the experimenter and the next trial would commence. The exposure to stimulus followed the 2 step up, 1 step down staircase approach as shown in figure \ref{fig:user_response}. So, the stimulus separation between audio and haptic for the out of sync feedback pair is determined based on user's previous input. For successive correct responses, the stimulus difference would be decreased, whereas a single incorrect response would lead to increase in stimulus latency between the audio haptic pair. This leads to each participant responding in ways where they move back and forth within a certain time window of stimulus separation. Each participant was continuously exposed in this manner until 8 reversal of responses was recorded. A representation of an actual participant response is shown in \ref{fig:user_response}.

We designed a within subjects study since we wanted to expose each participant to both latency conditions; audio preceding haptic and vice versa. Each participant underwent two experimental conditions; (i) negative audio haptic latency, where audio is played before the haptic feedback and (ii) positive audio haptic latency, where audio is played after the haptic feedback. To overcome ordering influence, each participant was randomly assigned to one of the experimental conditions to begin with and moved to the next one. Once the participants completed this stage, they were asked to complete a post task questionnaire to gather information on driving experiences, gender, age and other possible relevant information.

\subsection{Pilot Study}
We conducted pilot studies with participants that were employees of an automotive company. We used 4 participants to explore and establish certain parameters that were later applied to the user study. This included the starting stimulus offset for both haptic or audio first setup, number of reversal points, and the step size. We initially overestimated the starting stimulus offset for both audio and haptic first setup to be between 350 ms - 400 ms. However, it became quite obvious during the pilot studies that a more conservative estimation of approximately 260 ms was adequate. In addition, during the pilot studies, we used 9 reversal points to end the experiment for each session. However, we realized that, based on the length of the experiment and overall convergence of stimulus values, decreasing it to 8 reversals, excluding the first reversal in both cases, was more appropriate. In terms of step size, our initial estimate of 30ms as large or initial step size and 15 ms as the small step size didn’t seem to cause any issues and therefore, we continued with the same numbers for the user study. 
\subsection{Participants}
We recruited 15 external participants to conduct the user study out of which 7 were female and 8 were male. In terms of the age distribution, 7 were above the age of 40 and 8 below. In addition, all participants were experienced drivers, with 11 participants indicating that they had more than 10 years of driving experience, with only 2 participants indicating less than 5 years of driving experience. 3 participants indicated that they drove 4-6 times a week while the rest indicated that they drove daily. 
\subsection{Experimental Procedure}
\begin{figure}
\centering
\includegraphics[width = 0.25\linewidth]{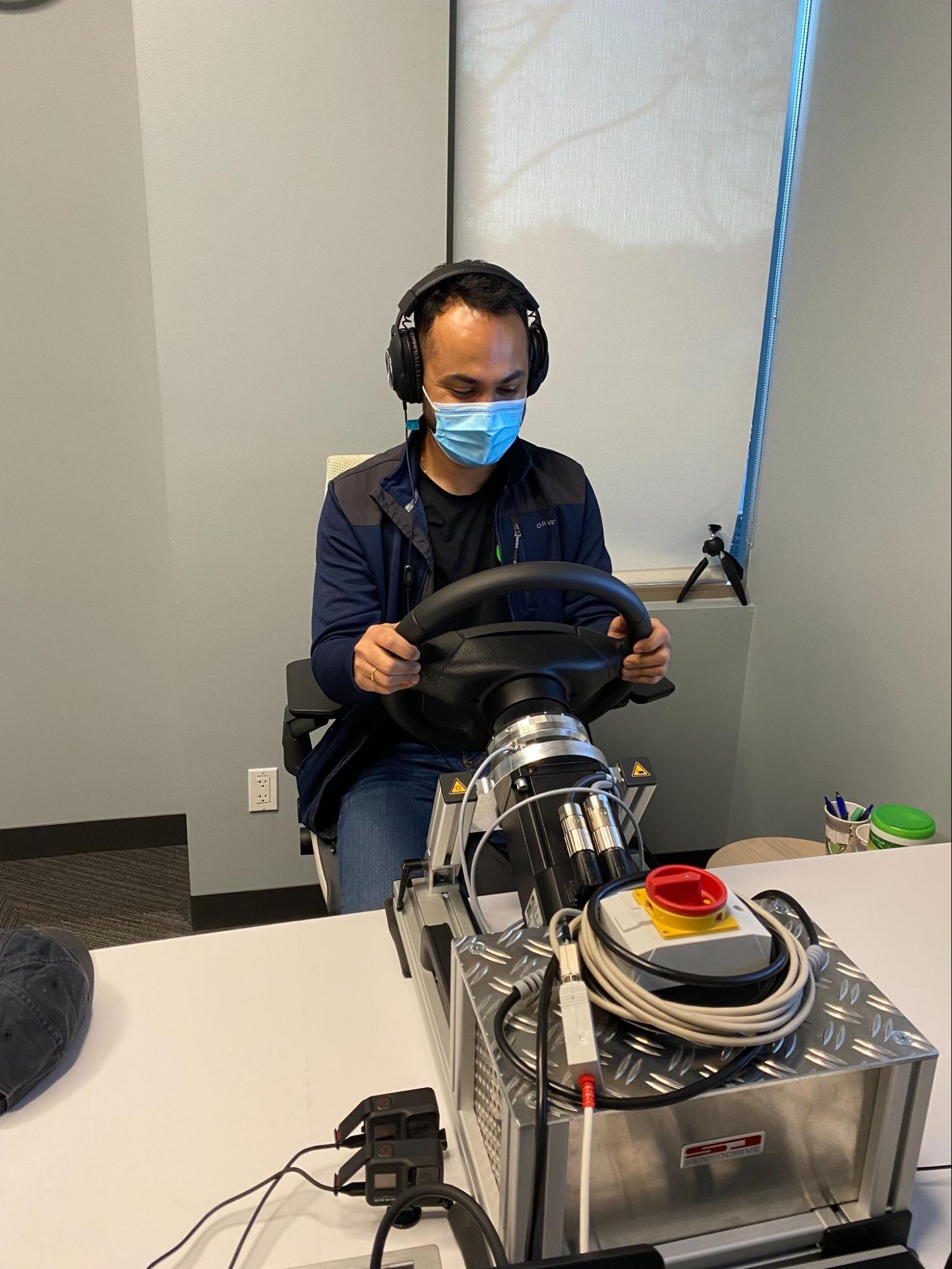}
\caption{Setup with the force feedback sensodrive wheel and headphone connected to the workstation. Not seen in the figure - Linux workstation.}
\label{fig:setup}
\end{figure}
All participants were hired using the services of \textit{Fieldwork San Francisco}. The entire experiment was conducted over a period of approximately one week. Once each participant arrived at the research facility, they went through all the covid protocols put in place before they were allowed inside. Each participant signed the consent form and also verbally consented to being recorded before they were given a brief description about the experiment and their task. Before commencing the experiment, each participant went through a simple test to make sure the sound being played on the wired headset was at an acceptable level while also making sure that they were able to feel the vibration on the steering wheel. The physical setup is shown in figure \ref{fig:setup}.

As a safety protocol, we instructed the participants to hold the steering wheel in a 10 - 2  position and make sure to not have their thumb or fingers inside the wheel to avoid injuries in an off chance that the wheel makes sudden movements. At this time, participants were allowed to ask any questions they might have before beginning the trials. After a session was concluded i.e 8 recorded reversals, participant were told that they could rest their hands for a minute before beginning the second session. Once this was over, they were instructed to fill out a post task questionnaire and this marked the end of the experimental session. Any thoughts, questions or concerns they had were verbally discussed and noted by the experimenter at this time before escorting them out of the research facility.
\section{Results}
\begin{figure}
\centering
\includegraphics[width = 0.99\linewidth]{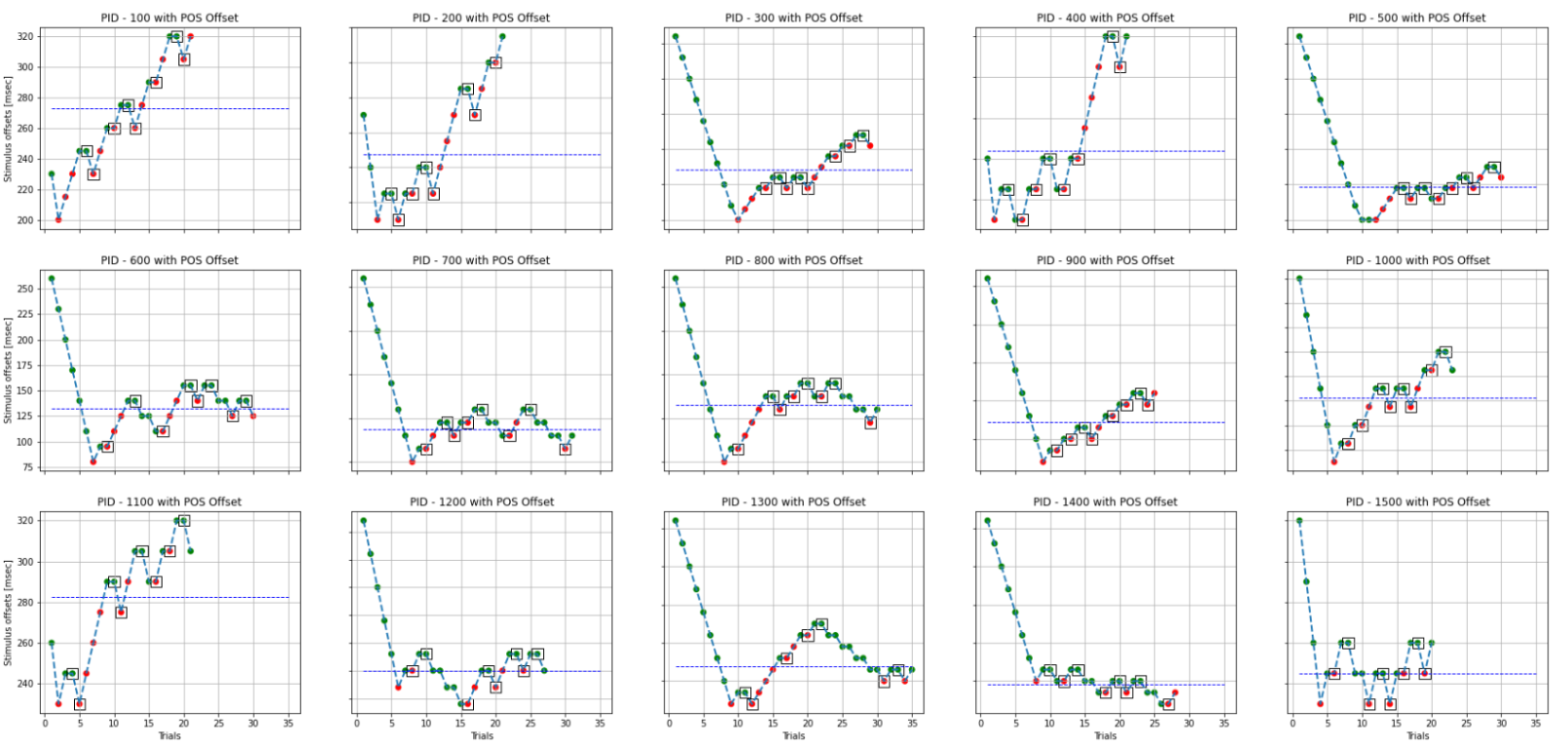}
\caption[]{Results of stimulus responses for all 15 participants in cases where haptic precedes audio.}
\label{fig:result1}
\end{figure}
\begin{figure}
\centering
\includegraphics[width = 0.99\linewidth]{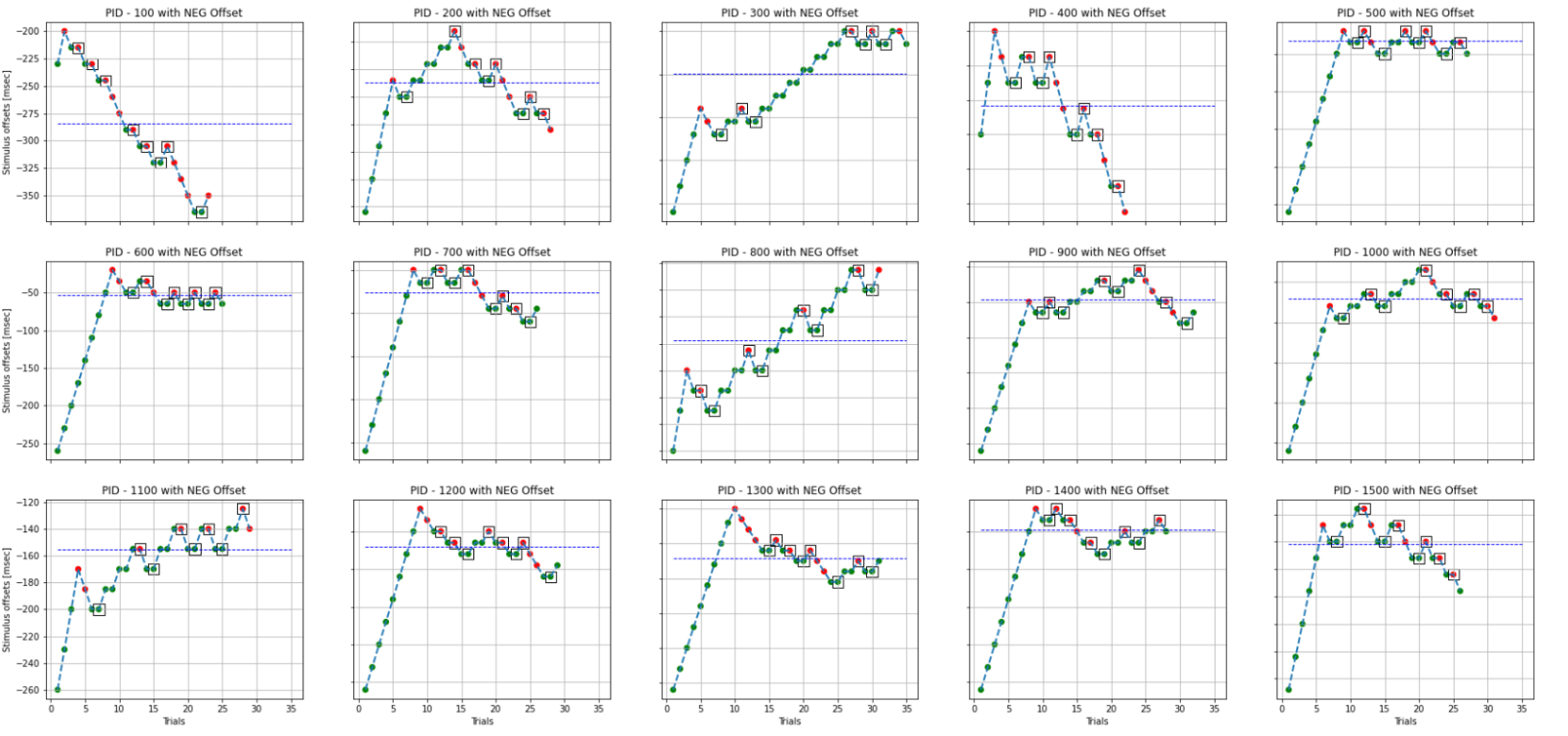}
\caption[]{Results of stimulus responses for all 15 participants in cases where audio precedes haptic.}
\label{fig:result2}
\end{figure}
\begin{figure*}
\begin{center}
\includegraphics[width = 0.99\linewidth]{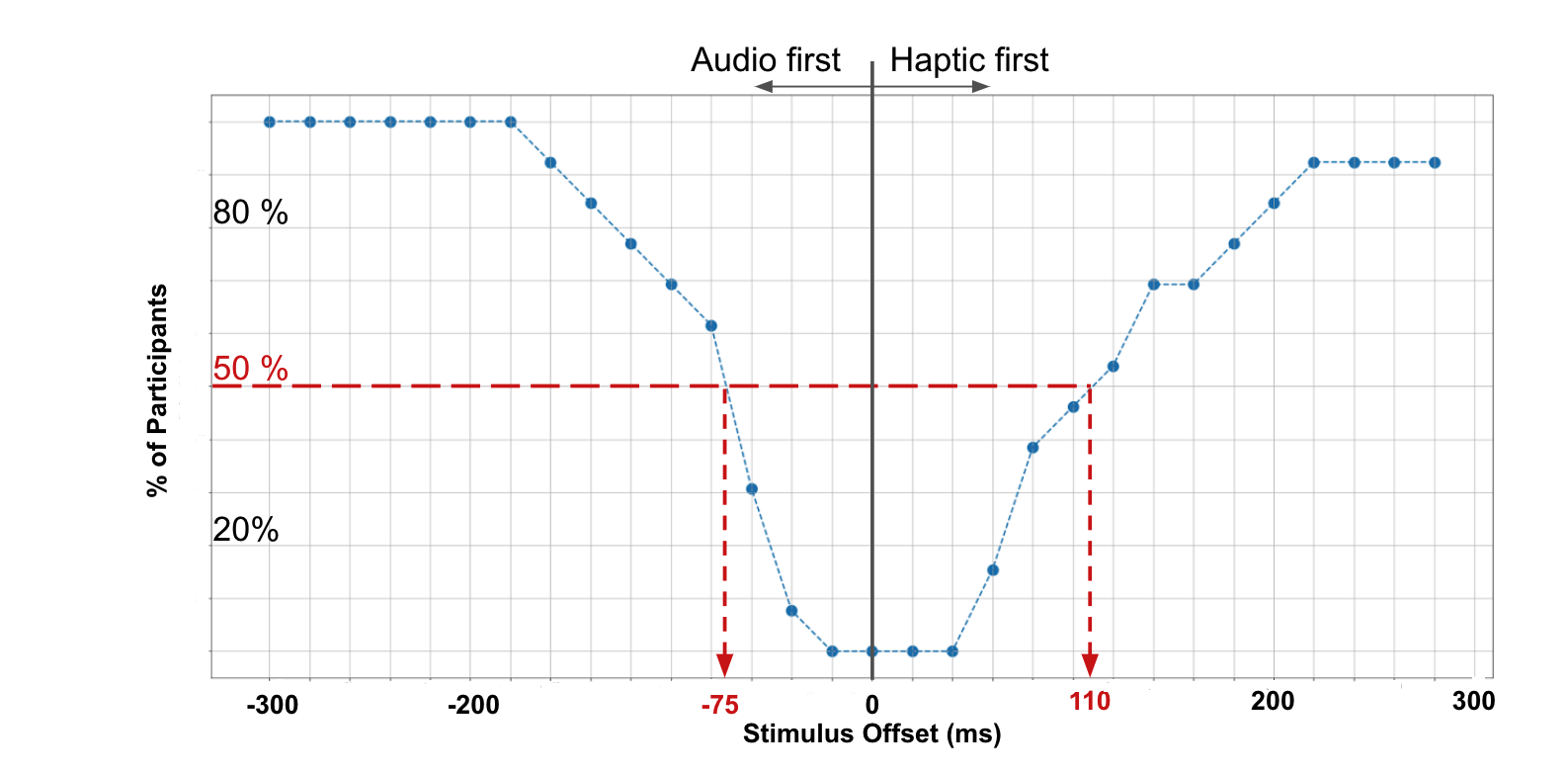}
\end{center}
\caption[]{Percentage of the participants who could detect the asynchrony. The red dotted line indicates the detection threshold for $50\%$ of participants.}
\label{fig:result}
\end{figure*}
The results for each participant is shown in figure \ref{fig:result1} and \ref{fig:result2}. In terms of the analysis of the overall results, participants 1 and 3 were discarded as outliers as their responses did not follow even the minimal expected pattern i.e they consistently failed to notice stimulus offsets that were larger than even the first stimulus value in both the audio first as well as the haptic first scenario. The DT was derived using equation \ref{eq:dt} for the remaining 13 participants. When haptic feedback preceded audio, the average DT was reported at $123.41 \pm 4.61$ ms and $87.73 \pm 4.61$ ms when audio preceded haptic. We were unable to observe any effects of age and gender in terms of the results.
\begin{equation}
    DT = \frac{1}{7}{\sum_{\alpha = 2}^{8} R_{\alpha}}
    \label{eq:dt}
\end{equation}

The cumulative score of participants based on their individual DTs is shown in figure \ref{fig:result}. This shows that in $50\%$ of cases, the detection threshold lies between -75 ms and 100 ms. Similarly, extrapolating from the same figure, we can also see that $20\%$ of the participants/drivers would notice a latency of 50 ms when audio is played before haptic and 65 ms when audio is played after haptic feedback. The implications based on these findings are briefly discussed below.
\section{Discussion}
The results indicate that the detection threshold for a haptic and audio based multi-modal feedback system in a driving context closely follow the findings from the past i.e audio preceding haptic is observable at smaller absolute time differences than the other way round. 

On the other hand, there are also design implications that can be derived based on the results reported above. For instance, safety systems in cars that rely on providing audio-haptic feedback, the robustness of the system can be tuned to accommodate the latency parameters based on our findings. In situations where, if for example, we want to design a feedback system that would only allow for $20\%$ of the participants to notice the asynchrony, latency window between -50 ms to 65 ms could be used, based on extrapolation on the graph shown in figure \ref{fig:result}. Similarly, an HMI interface that requires no more than 20\% of the drivers to perceive the asynchrony will have to optimize their computation power to include delivery of haptic and audio stimuli within the range of -50 ms to 65 ms. Based on the computational cost as well as the objective of the design, this can be tuned to either cover for more or less of the population. Our results provide a clear guideline in terms of what these human factors costs will amount to be.

\section{Limitations and Future Work}
This study was conducted entirely within a research setting and without the use of real driving situations or scenarios. This is a major limitation in regards to drawing accurate conclusions as the primary task for the participants in these scenarios was not driving. So, the DT estimates derived from this user study should be considered as conservative. We expect the DT for the same set of feedback; haptic and audio to be significantly higher if a similar setup were to be tested in a real driving context as cognitive resources are expended on higher priority tasks. However, to what extent such effects are characterized would be part of future research. The results presented in this study can be applied as baselines to draw meaningful extrapolations as needed.

The second major limitation was characterized by the Sensodrive wheel that was used as the primary component to provide haptic feedback. While the audio component is generated entirely in a digital format, that’s not exactly the case with the Sensodrive wheel which used the physical motors to generate the vibration. This in turn also creates sound artifacts that are clearly audible. The usage of over the ear noise canceling headphones mitigate this factor but does not entirely nullify the sound factor emanating from the wheel.

Overall, we were able to obtain results that we believe will have significant implications in designing multi-modal feedback systems as part of vehicle safety system framework. Our results, especially the one presented in figure \ref{fig:result}, can be applied to characterize system specifications when steering based multi-modal feedback systems are designed. Also, as evidenced by our results, more care can be put in place to make sure audio isn't played much earlier than other modalities as threshold of latency perception in such scenarios is lower.





\end{document}